\documentclass[runningheads]{llncs}
\usepackage{multirow}
\usepackage{booktabs}%
\usepackage{bm}
\usepackage{amsmath}
\usepackage{breqn}
\usepackage{anyfontsize} 
\usepackage{graphicx}
\usepackage{amsfonts}
\begin{document}
\title{A Multi-Task Deep Learning Framework to Localize the Eloquent Cortex in Brain Tumor Patients Using Dynamic Functional Connectivity}
%
%
\author{Naresh Nandakumar\inst{1},Niharika Shimona D'souza\inst{1} Komal Manzoor\inst{2}, Jay J. Pillai\inst{2}, Sachin K. Gujar\inst{2}, Haris I. Sair\inst{2}, and Archana Venkataraman\inst{1}}
\institute{Dept. of Electrical and Computer Engineering, Johns Hopkins University, USA \and Dept. of Neuroradiology, Johns Hopkins School of Medicine, USA
\email{nnandak1@jhu.edu}}
\authorrunning{N. Nandakumar et al. }
\titlerunning{Eloquent Cortex Localization with Multi-task Deep Learning}
\maketitle 
%
%
\begin{abstract}
We present a novel deep learning framework that uses dynamic functional connectivity to simultaneously localize the language and motor areas of the eloquent cortex in brain tumor patients. Our method leverages convolutional layers to extract graph-based features from the dynamic connectivity matrices and a long-short term memory (LSTM) attention network to weight the relevant time points during classification. The final stage of our model employs multi-task learning to identify different eloquent subsystems. Our unique training strategy finds a shared representation between the cognitive networks of interest, which enables us to handle missing patient data. We evaluate our method on resting-state fMRI data from 56 brain tumor patients while using task fMRI activations as surrogate ground-truth labels for training and testing. Our model achieves higher localization accuracies than conventional deep learning approaches and can identify bilateral language areas even when trained on left-hemisphere lateralized cases. Hence, our method may ultimately be useful for preoperative mapping in tumor patients.
\end{abstract}
\section{Introduction}
\par{}The eloquent cortex consists of regions in the brain that are responsible for language comprehension, speech, and motor function. Identifying and subsequently avoiding these areas during a neurosurgery is crucial for improving recovery and postoperative quality of life. However, localizing these networks is challenging due to the varying anatomical boundaries of the eloquent cortex across people \cite{ojemann1978language,tomasi2012language}. The language network has especially high interindividual variability because it can appear on one or both hemispheres \cite{tzourio2004interindividual}. The gold standard for preoperative functional mapping of eloquent areas is intraoperative electrocortical stimulation (ECS) of the cerebral cortex during surgery \cite{gupta2007awake,berger1989brain}. While reliable, ECS is highly invasive and requires the patient to be awake and responsive during surgery.
\par{}For these reasons, task-fMRI (t-fMRI) is becoming increasingly popular as a noninvasive alternative to ECS. Typically, activation maps derived from t-fMRI are inspected by an expert to determine the regions in the brain that are recruited during the experimental condition. However, t-fMRI can be unreliable for certain populations, like children and the cognitively disabled, due to their inability to complete the paradigm~\cite{kokkonen2009preoperative,lee2016clinical}. In contrast to t-fMRI, resting-state fMRI (rs-fMRI) captures spontaneous fluctuations in the brain when the subject is lying passively in the scanner. Unlike t-fMRI paradigms, which are designed to activate a single area, correlations in the rs-fMRI data can be used to identify multiple cognitive systems~\cite{van2010exploring}. Recent work has moved towards using rs-fMRI for presurgical mapping to avoid the above issues associated with t-fMRI~\cite{lee2016clinical,nandakumar2019novel}.
\par{}Automatically localizing the eloquent cortex using rs-fMRI is a challenging problem with limited success in the literature. For example, the authors of \cite{sair2016presurgical,tie2014defining} demonstrate that spatial components identified by group ICA on the rs-fMRI data coincide with the language and motor networks from t-fMRI. While the result is promising, the spatial accuracy is highly variable across patients. The work in \cite{hacker2013resting} describes a multi-layer perceptron architecture that classifies rs-fMRI networks at the voxel level using seed based correlation maps; this method was extended in \cite{leuthardt2018integration} to handle tumor cases. However, the perceptron is trained on healthy subjects and may not accommodate changes in brain organization due to the tumor. Finally, the method in \cite{nandakumar2019novel} is the first end-to-end graph neural network (GNN) to automatically localize eloquent cortex in tumor patients. While this method achieves good classification performance, separate GNNs must be trained and tested for each eloquent network \cite{nandakumar2019novel}, which increases the training time, the overall number of parameters, and the required training data.
\par{}There is growing evidence in the field that functional connectivity patterns are not static, but evolve over time. In particular, studies have shown that individual functional systems are more strongly present during specific intervals of the rs-fMRI scan~\cite{dvornek2019jointly,yan2018deep}. Several studies have leveraged these dynamic connectivity patterns for classification. For example, the work in \cite{dvornek2017identifying} uses a long-short term memory (LSTM) cell to learn time dependencies within the rs-fMRI to discriminate patients with autism from controls. More recent work by \cite{rashid2016classification} and \cite{el2019hybrid} has shown that combining static and dynamic connectivity can achieve better patient versus control classification performance than either set of features alone. However, these works focus on group-level discrimination. We will leverage similar principles in this paper to classify ROIs within a single~patient. 
\par{}We propose a novel multi-task deep learning framework that uses both convolutional nerual networks (CNNs) and an LSTM attention network to extract and combine dynamic connectivity features for eloquent cortex localization. The final stage of our model employs multi-task learning (MTL) to implicitly select the relevant time points for each network and simultaneously identify regions of the brain involved in language processing and motor functionality. Our model finds a shared representation between the cognitive networks of interest, which enables us to handle missing data. This coupling also reduces the number of model parameters, so that we can learn from limited patient data. We evaluate our framework on rs-fMRI data from 56 brain tumor patients while using task fMRI activations as surrogate ground-truth labels for training and testing. Our model achieves higher localization accuracies than a variety of baseline techniques, thus demonstrating its promise for preoperative mapping.
\section{Eloquent Cortex Localization Using Deep Learning}
\begin{figure*}[t!]
\begin{centering}
\includegraphics[width=.9\textwidth,height=\textheight,keepaspectratio]{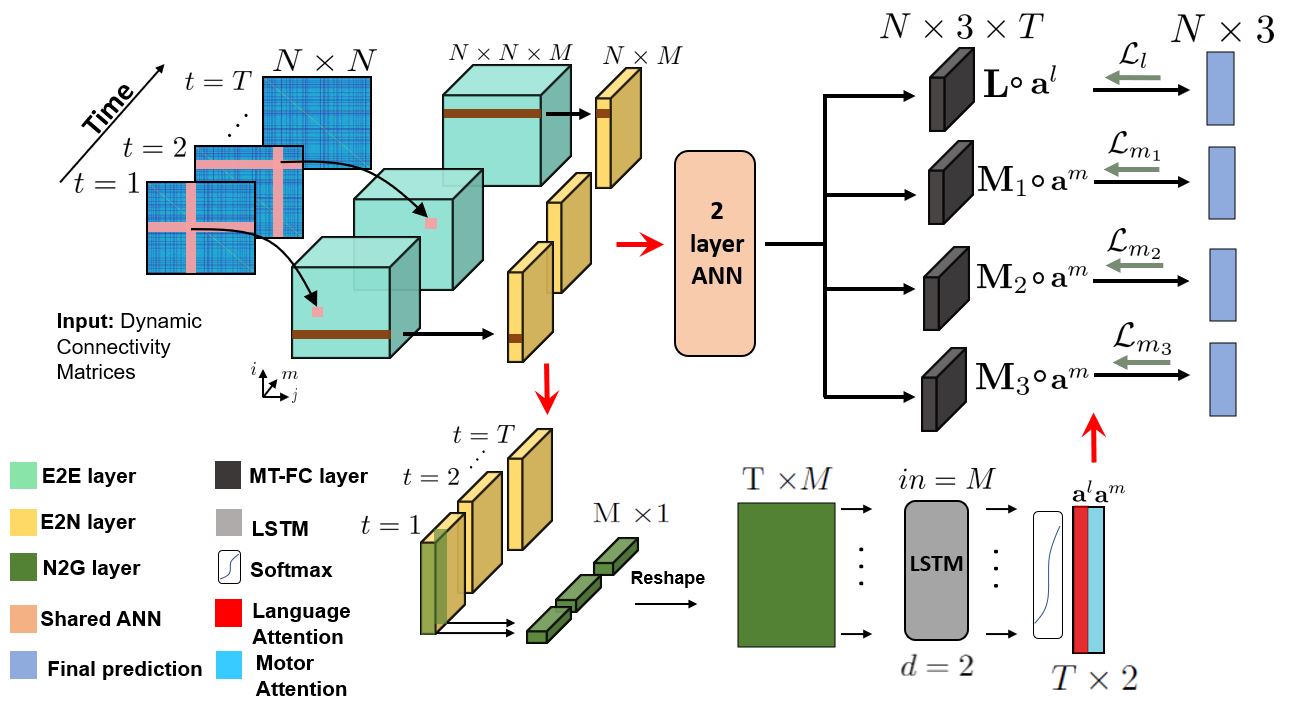}
\caption{\textbf{Top}: Specialized convolutional layers identify dynamic patterns that are shared across the functional systems. \textbf{Bottom}: The dynamic features are input to an LSTM network to learn attention weights $\mathbf{a}^l$ (language) and $\mathbf{a}^m$ (motor). \textbf{Right}: MTL to classify the language ($\mathbf{L}$), finger ($\mathbf{M}_1$), foot ($\mathbf{M}_2$) and tongue ($\mathbf{M}_3$) networks.}
\end{centering}
\end{figure*}
Our framework makes two underlying assumptions. First, while the anatomical boundaries of the eloquent cortex may shift across individuals, its functional connectivity with the rest of the brain will be preserved \cite{sair2016presurgical}. Second, the networks associated with the eloquent cortex phase in and out of synchrony across the rs-fMRI scan \cite{kunert2019extracting}. Hence, isolating these key time points will help to refine our localization. Fig. 1 illustrates our framework. In the top branch, we use specialized convolutional filters to capture rs-fMRI co-activation patterns from the dynamic connectivity matrices. In the bottom branch, we use an LSTM to identify key time points where the language and/or motor networks are more synchronous. We tie the activations from the LSTM branch of our model into our MTL classification problem via our specialized loss function.
\paragraph{\textbf{Input Connectivity Matrices.}}
We use the sliding window technique to obtain our connectivity matrices~\cite{hutchison2013dynamic}. Let $N$ be the number of brain regions in our parcellation, $T$ be the total number of sliding windows (i.e., time points in our model), and $\{\mathbf{W}^t\}_{t=1}^{T} \in \mathbb{R}^{N \times N}$ be the dynamic similarity matrices. $\mathbf{W}^t$ is constructed from the input time courses $\{\mathbf{X}^t\}_{t=1}^{T} \in \mathbb{R}^{D \times N}$, where each $\mathbf{X}^{t}$ is a segment of the rs-fMRI obtained with window size $D$. The input $\mathbf{W}^{t} \in \mathbb{R} ^{N \times N}$~is
\begin{equation}
\mathbf{W}^{t} = \exp \Bigg[ \frac{(\mathbf{X}^t)^T \mathbf{X}^t}{\epsilon} - 1 \Bigg]
\end{equation}
where $\epsilon \geq 1$ is a user-specified parameter that controls decay speed~\cite{nandakumar2019novel}. Recall that our setup must accommodate the presence of brain tumors that vary across patients. Since these tumors represent non-functioning areas of the brain, we follow the approach of \cite{nandakumar2019novel} and treat the corresponding rows and columns of the simlarity matrix as ``missing data" by fixing them to zero. This procedure removes the effect of the tumor regions on the downstream convolution operations. 
\paragraph{\textbf{Representation Learning for Dynamic Connectivity.}} Our network leverages the specialized convolutional layers developed in \cite{kawahara2017brainnetcnn} for static analysis. The edge-to-edge (E2E) layer in Fig.1 acts across rows and columns of the input matrix $\mathbf{W}^t$. Mathematically, let $f~\in~\{1,\cdots,F\}$ be the E2E filter index, $\mathbf{r}^{f} \in \mathbb{R}^{1 \times N}$ be the row filter $f$, $\mathbf{c}^{f} \in \mathbb{R}^{N \times 1}$ be the column filter $f$,  $\mathbf{b} \in \mathbb{R}^{F \times 1}$ be the E2E bias, and $\bm{\phi}(.)$ be the activation function. For each time point $t$ the feature map $\mathbf{H}^{f,t} \in \mathbb{R}^{N \times N}$ is computed as follows:
\begin{equation}
\mathbf{H}^{f,t}_{i,j} = \bm{\phi} \Bigg( \sum_{n=1}^{N}\mathbf{r}^f_n \mathbf{W}^{t}_{i,n} + \mathbf{c}^f_n \mathbf{W}^{t}_{n,j} + \mathbf{b}_{f} \Bigg).\label{eq3}
\end{equation}
Effectively, the E2E filter output $\mathbf{H}_{ij}^{f,t}$ for edge $(i,j)$ extracts patterns associated with the neighborhood connectivity of node $i$ and node $j$. The edge-to-node (E2N) filter in Fig. 1 is a 1D convolution along the columns of each feature map. Mathematically, let $\mathbf{g}^{f} \in \mathbb{R}^{N \times 1}$ be E2N filter $f$ and $\mathbf{p} \in \mathbb{R}^{F \times 1}$ be the E2N bias. The E2N output $\mathbf{h}^{f,t} \in \mathbb{R}^{N \times 1}$ from input $\mathbf{H}^{f,t}$ is computed as
\begin{equation}
\mathbf{h}^{f,t}_{i} =  \bm{\phi} \Bigg( \sum_{n=1}^{N} \mathbf{g}^{f}_{n} \mathbf{H}^{f,t}_{i,n}  +  \mathbf{p}_{f} \Bigg). \label{eq4}
\end{equation}
The E2E and E2N layers extract topological graph-theoretic features from the connectivity data. Following the convolutional layers in the top branch, we cascade two fully-connected (FC) layers to combine these learned topological features for our downstream multi-task classification. In the bottom branch, we use a node-to-graph (N2G) layer to extract features that will be input to our LSTM network. The N2G filter acts as a 1D convolution along the first dimension of the E2N output, effectively collapsing the node information to a low dimensional representation for each time point. Let $\mathbf{k}^{f} \in \mathbb{R}^{N \times 1}$ be N2G filter $f$ and $\mathbf{d} \in \mathbb{R}^{F \times 1}$ be the bias. The N2G filter gives a scalar output $q^{f,t}$ for each input $\mathbf{h}^{f,t}$ by 
\begin{equation}
q^{f,t} = \bm{\phi} \Bigg( \sum_{n=1}^{N} \mathbf{k}_n^{f}\cdot \mathbf{h}_n^{f,t} + \mathbf{d}_f \Bigg) .
\end{equation}
\paragraph{\textbf{Dynamic Attention Model.}} Per time point, we define $\mathbf{q}^t=[q^{1,t} \cdots q^{F,t}]$ and feed the vectors $\{\mathbf{q}^t\}_{t=1}^{T}$ into an LSTM module to learn attention weights for our classification problem. The LSTM adds a cell state to the basic recurrent neural network to help alleviate the vanishing gradient problem, essentially by accumulating state information over time \cite{xingjian2015convolutional}. LSTMs have demonstrated both predictive power for rs-fMRI analysis \cite{dvornek2017identifying,el2019hybrid} and the ability to identify different brain states \cite{li2018brain}. We choose $d=2$ as the output dimension, and perform a softmax over each column of the LSTM output to get the attention vectors $\mathbf{a}^l \in \mathbb{R}^{T \times 1}$ (language) and $\mathbf{a}^m \in \mathbb{R}^{T \times 1}$ (motor). These attention vectors provide information on which input connectivity matrices are more informative for identifying the language or motor networks. The attention model outputs are combined with the classifer during backpropogation in our novel loss function.
\paragraph{\textbf{Multi-task Learning with Incomplete Data.}}The black blocks in Fig. 1 show the multi-task FC (MT-FC) layers, where we have four separate branches to identify the language, finger, foot, and tongue areas. Up until this point, there has been an entirely shared representation of the feature weights at each layer. Let $\mathbf{L}^t,\mathbf{M}^t_1,\mathbf{M}^t_2,$ and $\mathbf{M}^t_3 \in \mathbb{R}^{N \times 3}$ be the output of the language, finger, foot, and tongue MT-FC layers, respectively, at time $t$. The $N \times 3$ matrix represents the region-wise assignment into one of three classes; eloquent, tumor, and background. As in~\cite{nandakumar2019novel}, we introduce the tumor as its own learned class to remove any bias these regions may have introduced to the algorithm.
\par{} We introduce a novel variant of a modified version of the risk-sensitive cross-entropy loss function~\cite{suresh2008risk,nandakumar2019novel}, which is designed to handle membership imbalance in multi-class problems. Let $\delta_c$ be the risk factor associated with class $c$. If $\delta_c$ is small, then we pay a smaller penalty for misclassifying samples that belong to class $c$. Due to a training set imbalance, we set different values for the language class ($\delta^{l}_c$) and motor classes ($\delta^{m}_c$) respectively. Let $\mathbf{Y}^{l}, \mathbf{Y}^{m_1},  \mathbf{Y}^{m_1},$ and $ \mathbf{Y}^{m_3} \in \mathbb{R} ^ {N \times 3}$ be one-hot encoding matrices for the ground-truth class labels of the language and motor subnetworks. Notice that our framework allows for overlapping eloquent labels, as brain regions can be involved in multiple cognitive processes. Our loss function is the sum of four terms:
\begin{equation}
\begin{split}
\mathcal{L}_{\Theta}(\{\mathbf{W}^t\}_{t=1}^{T},\mathbf{Y}) =   \sum_{n=1}^{N} \sum_{c=1}^{3} \Big[ \underbrace{- \delta^{l}_c   \log \Big( \sigma   \Big(\sum_{t=1}^{T}\mathbf{L}^t_{n,c} \cdot \mathbf{a}^{l,t}\Big)\Big)\mathbf{Y}^{l}_{n,c}}_{\text{Language Loss } \mathcal{L}_l} \\
 \underbrace{- \delta^{m}_c   \log \Big(\sigma \Big(\sum_{t=1}^{T}\mathbf{M}^t_{1n,c} \cdot \mathbf{a}^{m,t}\Big)\Big)\mathbf{Y}^{m_1}_{n,c}}_{\text{Finger Loss } \mathcal{L}_{m_1}}  \underbrace{- \delta^{m}_c   \log \Big(\sigma \Big(\sum_{t=1}^{T}\mathbf{M}^t_{2n,c} \cdot \mathbf{a}^{m,t}\Big)\Big)\mathbf{Y}^{m_2}_{n,c}}_{\text{Foot Loss } \mathcal{L}_{m_2}} \\  \underbrace{- \delta^{m}_c   \log \Big(\sigma \Big(\sum_{t=1}^{T}\mathbf{M}^t_{3n,c} \cdot \mathbf{a}^{m,t}\Big)\Big)\mathbf{Y}^{m_3}_{n,c}}_{\text{Tongue Loss } \mathcal{L}_{m_3}}\Big]  
\end{split}
\end{equation}
where $\sigma(\cdot)$ is the sigmoid function. Our loss in Eq. (5) allows us to handle missing information during training. For example, if we only have ground-truth labels for some of the functional systems, then we can freeze the other branches and just backpropagate the known loss terms. This partial backpropagation will continue to refine the shared representation, thus maximizing the amount of information mined from our training data. Note that our formulation is agnostic to the length of the rs-fMRI scan (i.e. $T$), which is useful in clinical practice.
\paragraph{\textbf{Implementation details.}}
We implement our network in PyTorch using the SGD optimizer with weight decay $=5 \times 10^{-5}$ for parameter stability, and momentum $=0.9$ to improve convergence. We train our model with learning rate $=0.002$ and $300$ epochs, which provides for reliable performance without overfitting. We used $D=45$ and a stride length of $5$ for the sliding window. We specified $F=25$ feature maps in the convolutional branch, and $2$ layers in our LSTM. The LeakyReLU with slope $=-0.1$ was used for $\bm{\phi}(.)$. Using cross validation, we set the cross-entropy weights to $\bm{\delta}^m =(1.5,0.5,0.2)$, and $\bm{\delta}^l =(2.25,0.5,0.2)$. 
\par{}We compare the performance of our model against three baselines:
\begin{itemize}
\item[1.]{PCA + Multi-class linear SVM on dynamic connectivity matrices (SVM)}
\item[2.]{A multi-task GNN on static connectivity (MT-GNN)}
\item[3.]{A multi-task ANN with LSTM attention model (MT-ANN)}
\end{itemize}
The first baseline is a traditional machine learning SVM approach to our problem. The MT-GNN operates on static connectivity and does not have an LSTM module. We include the MT-GNN to observe the difference in performance with and without using dynamic information. The MT-ANN maintains the same number of parameters as our model but has fully-connected layers instead of convolutional layers. Therefore, the MT-ANN does not consider the network organization of the input dynamic connectivity matrices.
\section{Experimental Results}
\begin{figure*}[t!]
\begin{centering}
\includegraphics[width=.75\textwidth,height=\textheight,keepaspectratio]{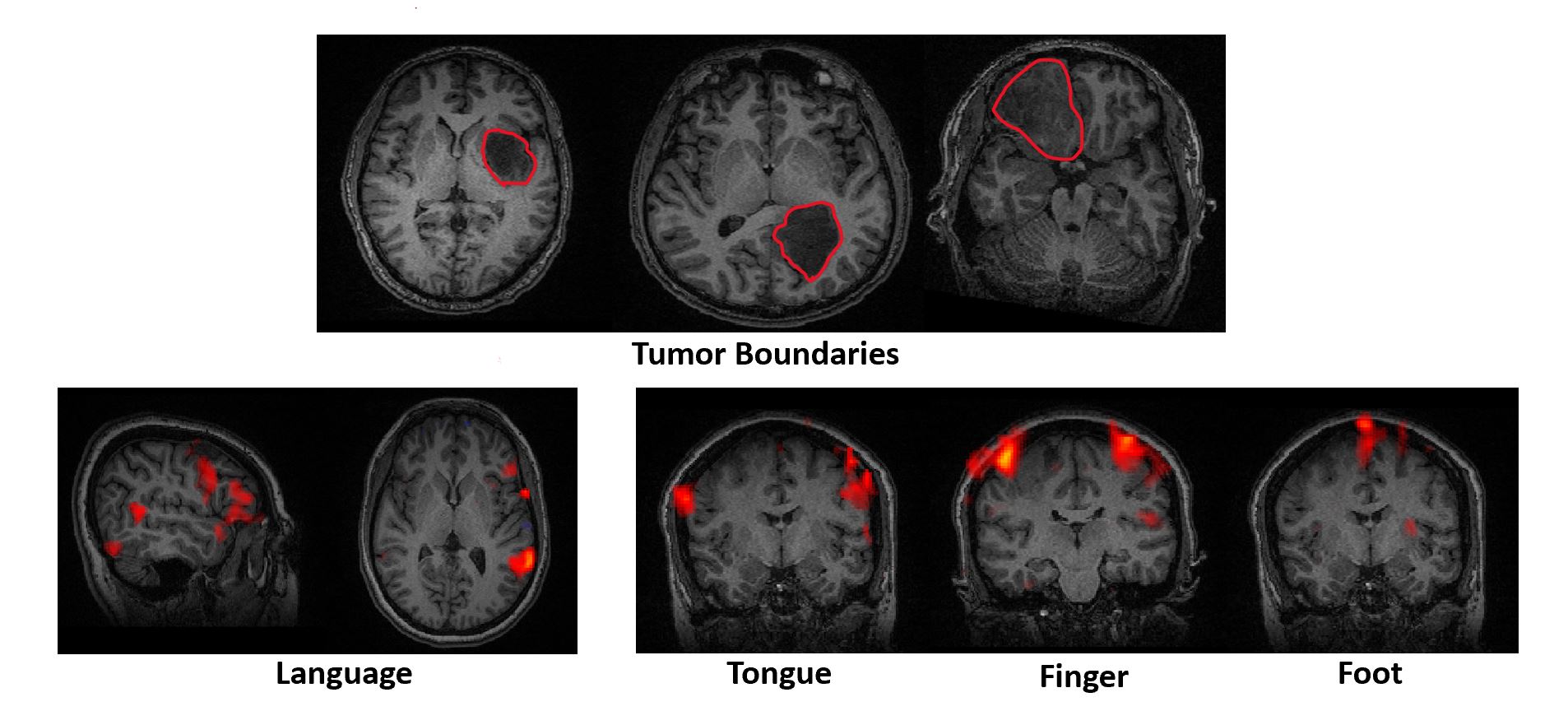}
\caption{\textbf{Top}: Tumor boundaries for three patients. \textbf{Left}: One sagital and axial view of a language network. \textbf{Right}: Coronal views of the motor sub-networks for one patient.}
\end{centering}
\end{figure*}
\paragraph{\textbf{Dataset and Preprocessing.}}
We evaluate the methods on rs-fMRI data from $56$ brain tumor patients who underwent preoperative mapping at our institution. These patients also underwent t-fMRI scanning, which we use to derive pseudo ground-truth labels for training and validation. Our dataset includes three different motor paradigms that are designed to target distinct parts of the motor homunculus~\cite{jack1994sensory}: finger tapping, tongue moving, and foot tapping. It also includes two language paradigms, sentence completion and silent word generation. Since the t-fMRI data was acquired for clinical purposes, not all patients performed each task. The number of subjects that performed the language, finger, foot, and tongue tasks are displayed in the left column of Table~1. 
\par{}The fMRI data was acquired using a $3.0$ T Siemens Trio Tim (TR $= 2000$ ms, TE $=30$ ms, FOV $=24$ cm, res $=3.59 \times 3.59 \times 5$ mm). Preprocessing steps include slice timing correction, motion correction and registration to the MNI-152 template. The rs-fMRI was further bandpass filtered from $0.01$ to $0.1$ Hz, spatially smoothed with a $6$ mm FWHM Gaussian kernel, scrubbed using the ArtRepair toolbox \cite{mazaika2009methods} in SPM8, linearly detrended, and underwent nuisance regression using the CompCor package \cite{behzadi2007component}. We used the Craddocks atlas to obtain $N$=384 brain regions \cite{craddock2012whole}. Tumor boundaries for each patient were manually delineated by a medical fellow using the MIPAV software package \cite{mcauliffe2001medical}. An ROI was determined as belonging to the eloquent class if a majority of its voxel membership coincided with that of the t-fMRI activation map. Tumor labels were determined in a similar fashion according to the MIPAV segmentations. A general linear model implemented in SPM8 was used to obtain t-fMRI activation maps. Fig. 2 shows representative examples of the tumor boundaries and each of the four cognitive networks of interest obtained from t-fMRI. 
\begin{table}[t!]

\caption{Class accuracy, overall accuracy, and ROC statistics. The number in the first column indicates number of patients who performed the task.}\label{tab1}
\begingroup
\setlength{\tabcolsep}{8.5pt} 
\renewcommand{\arraystretch}{1}
\begin{center}
 \begin{tabular}{ c c c c c} 
 \hline 
Task &Method & Eloquent     & Overall   & AUC\\
\hline
Language (56) & SVM  & $0.49$ &  $0.59$ & $0.55$\\
 & MT-ANN  & $0.70$ &  $0.71$ & $0.70$\\
 & MT-GNN &   $0.73$ & $0.74$ & $0.74$\\
 & Proposed  &  $\mathbf{0.85}$ & $\mathbf{0.81}$ & $\mathbf{0.80}$\\
\hline
Finger (36) & SVM  & $0.54$ & $0.61$ & $0.57$ \\
 & MT-ANN  & $0.73$ & $0.75$ & $0.74$ \\
 & MT-GNN &  $\underline{0.87}$ & $\mathbf{0.86}$ & $\mathbf{0.84}$ \\
 & Proposed  & $\mathbf{0.88}$ & $\underline{0.85}$ & $\mathbf{0.84}$ \\

\hline
Foot (17) &  SVM  & $0.58$ & $0.63$ & $0.60$ \\
 & MT-ANN  & $0.72$ & $0.77$ & $0.74$\\
 & MT-GNN &  $0.82$ & $0.79$ &  $0.79$ \\ 
 & Proposed  & $\mathbf{0.86}$ & $\mathbf{0.85}$ & $\mathbf{0.82}$ \\

\hline
Tongue (39) & SVM  & $0.54$ & $0.60$ & $0.58$ \\
 & MT-ANN  & $0.74$ & $0.76$ & $0.73$ \\
 & MT-GNN &  $0.85$ & $0.81$ &  $0.82$ \\
 & Proposed  & $\mathbf{0.87}$ & $\mathbf{0.83}$ & $\mathbf{0.84}$\\
\hline
\end{tabular}
\end{center}
\endgroup
\end{table}

\paragraph{\textbf{Localization.}}
We use 8-fold cross validation (CV) to quantify our eloquent cortex localization performance. Table 2 reports the eloquent per-class accuracy and the area under the receiver operating characteristic curve (AUC) for detecting the eloquent class on the testing data. Each MT-FC branch has separate metrics. Our proposed method has the best overall performance, as highlighted in bold. Even with attention from the LSTM layer, we observe that a fully-connected ANN still is sub-par for our task compared to using the specialized E2E, E2N, and N2G layers. Furthermore, our performance gains are most notable when classifying the language and foot networks. The former is particularly relevant for preoperative mapping, due to the difficulties in identifying the language network even with ECS \cite{ojemann1978language,tomasi2012language}. 
Figure 3 shows the language (left) and motor (right) attention vectors for all patients across time. We observe that both systems phase in and out, such that when one system is more active, the other is less active. This pattern lends credence to our hypothesis that identifying the critical intervals is key for localization. Hence, our model outperforms the static MT-GNN.

\begin{figure*}[t!]
\begin{centering}
\includegraphics[width=.8\textwidth,height=\textheight,keepaspectratio]{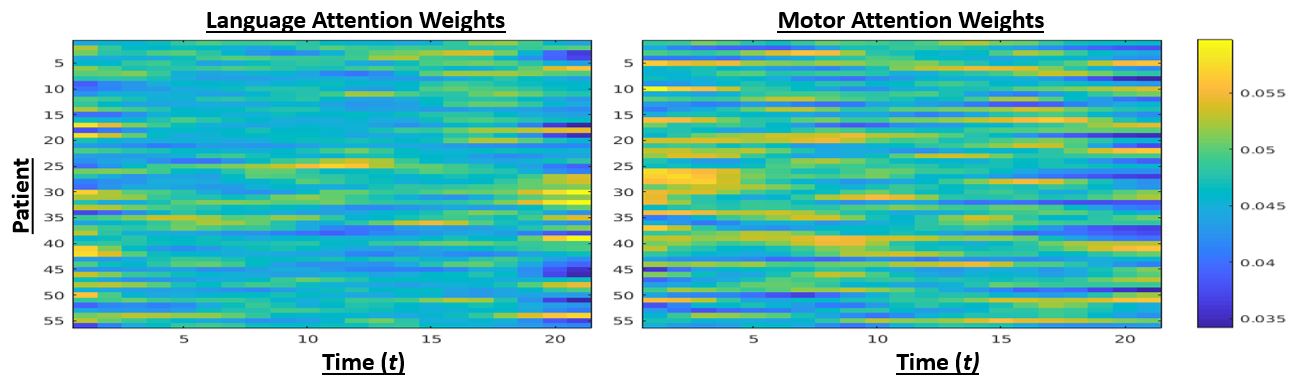}
\caption{Language (\textbf{L}) and motor (\textbf{R}) attention weights for all patients. }
\end{centering}
\end{figure*}
\paragraph{\textbf{Bilateral Language Identification.}}
Finally, we test whether our model can recover a bilateral language network, even when this case is not present in the training data. Here, we trained the model on $51$ left-hemisphere language network patients and tested on the remaining $5$ bilateral patients. Our model correctly predicted bilateral parcels in all five subjects. Fig. 4 shows ground truth (blue) and predicted language maps (yellow) for two example cases. The mean language class accuracy for these five cases was $\mathbf{0.72}$. This is slighly lower than reported in Table 1 likely due to the mismatch in training information.
\begin{figure*}[t!]
\begin{centering}
\includegraphics[width=.7\textwidth,height=\textheight,keepaspectratio]{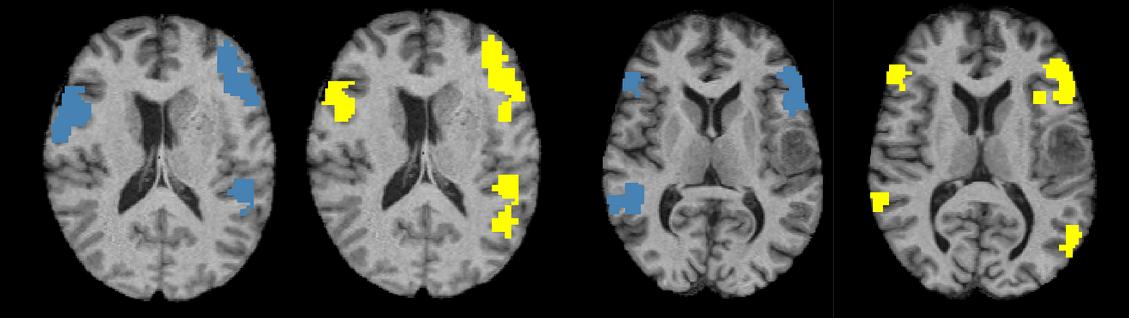}
\caption{Ground truth (Blue) and predicted (Yellow) language labels for two subjects. }
\end{centering}
\end{figure*}

\section{Conclusion}

We have demonstrated a novel multi-task learning framework that uses dynamic functional connectivity to identify separate sub-systems of the eloquent cortex in brain tumor patients. Our model is extendable to adding more eloquent sub-classes, as it finds a shared representation of the eloquent cortex that can subsequently classify sub-regions of interest. Going one step further, we show that our model can correctly identify bilateral language networks even when trained on only unilateral cases. Finally, our attention features suggest that using dynamic connectivity could be preferred to the traditional static case. Our results demonstrate promise for using rs-fMRI analysis in the preoperative phase for tumor resection procedures. \\ \\ \textbf{Acknowledgements:} This work was supported by the National Science Foundation CAREER award 1845430 (PI: Venkataraman) and the Research \& Education Foundation Carestream Health RSNA Research Scholar Grant RSCH1420.
\bibliographystyle{ieeetr}
\bibliography{MyRefs}

\begin{thebibliography}{10}

\bibitem{ojemann1978language}
G.~A. Ojemann and H.~A. Whitaker, ``Language localization and variability,''
  {\em Brain and language}, vol.~6, no.~2, pp.~239--260, 1978.

\bibitem{tomasi2012language}
D.~Tomasi and N.~Volkow, ``Language network: segregation, laterality and
  connectivity,'' {\em Molecular psychiatry}, vol.~17, no.~8, p.~759, 2012.

\bibitem{tzourio2004interindividual}
N.~Tzourio-Mazoyer, G.~Josse, F.~Crivello, and B.~Mazoyer, ``Interindividual
  variability in the hemispheric organization for speech,'' {\em Neuroimage},
  vol.~21, no.~1, pp.~422--435, 2004.

\bibitem{gupta2007awake}
D.~K. Gupta, P.~Chandra, B.~Ojha, B.~Sharma, A.~Mahapatra, and V.~Mehta,
  ``Awake craniotomy versus surgery under general anesthesia for resection of
  intrinsic lesions of eloquent cortex—a prospective randomised study,'' {\em
  Clinical neurology and neurosurgery}, vol.~109, no.~4, pp.~335--343, 2007.

\bibitem{berger1989brain}
M.~S. Berger, J.~Kincaid, G.~A. Ojemann, and E.~Lettich, ``Brain mapping
  techniques to maximize resection, safety, and seizure control in children
  with brain tumors,'' {\em Neurosurgery}, vol.~25, no.~5, pp.~786--792, 1989.

\bibitem{kokkonen2009preoperative}
S.-M. Kokkonen, J.~Nikkinen, J.~Remes, J.~Kantola, T.~Starck, M.~Haapea,
  J.~Tuominen, O.~Tervonen, and V.~Kiviniemi, ``Preoperative localization of
  the sensorimotor area using independent component analysis of resting-state
  fmri,'' {\em Magnetic resonance imaging}, vol.~27, no.~6, pp.~733--740, 2009.

\bibitem{lee2016clinical}
M.~H. Lee, M.~M. Miller-Thomas, T.~L. Benzinger, D.~S. Marcus, C.~D. Hacker,
  E.~C. Leuthardt, and J.~S. Shimony, ``Clinical resting-state fmri in the
  preoperative setting: are we ready for prime time?,'' {\em Topics in magnetic
  resonance imaging: TMRI}, vol.~25, no.~1, p.~11, 2016.

\bibitem{van2010exploring}
M.~P. Van Den~Heuvel and H.~E.~H. Pol, ``Exploring the brain network: a review
  on resting-state fmri functional connectivity,'' {\em European
  neuropsychopharmacology}, vol.~20, no.~8, pp.~519--534, 2010.

\bibitem{nandakumar2019novel}
N.~Nandakumar, K.~Manzoor, J.~J. Pillai, S.~K. Gujar, H.~I. Sair, and
  A.~Venkataraman, ``A novel graph neural network to localize eloquent cortex
  in brain tumor patients from resting-state fmri connectivity,'' in {\em
  International Workshop on Connectomics in Neuroimaging}, pp.~10--20,
  Springer, 2019.

\bibitem{sair2016presurgical}
H.~I. Sair {\em et~al.}, ``Presurgical brain mapping of the language network in
  patients with brain tumors using resting-state fmri: Comparison with task
  fmri,'' {\em Human brain mapping}, vol.~37, no.~3, pp.~913--923, 2016.

\bibitem{tie2014defining}
Y.~Tie {\em et~al.}, ``Defining language networks from resting-state fmri for
  surgical planning—a feasibility study,'' {\em Human brain mapping},
  vol.~35, no.~3, pp.~1018--1030, 2014.

\bibitem{hacker2013resting}
C.~D. Hacker, T.~O. Laumann, N.~P. Szrama, A.~Baldassarre, A.~Z. Snyder, E.~C.
  Leuthardt, and M.~Corbetta, ``Resting state network estimation in individual
  subjects,'' {\em Neuroimage}, vol.~82, pp.~616--633, 2013.

\bibitem{leuthardt2018integration}
E.~C. Leuthardt, G.~Guzman, S.~K. Bandt, C.~Hacker, A.~K. Vellimana,
  D.~Limbrick, M.~Milchenko, P.~Lamontagne, B.~Speidel, J.~Roland, {\em
  et~al.}, ``Integration of resting state functional mri into clinical
  practice-a large single institution experience,'' {\em PloS one}, vol.~13,
  no.~6, p.~e0198349, 2018.

\bibitem{dvornek2019jointly}
N.~C. Dvornek, X.~Li, J.~Zhuang, and J.~S. Duncan, ``Jointly discriminative and
  generative recurrent neural networks for learning from fmri,'' in {\em
  International Workshop on Machine Learning in Medical Imaging}, pp.~382--390,
  Springer, 2019.

\bibitem{yan2018deep}
W.~Yan, H.~Zhang, J.~Sui, and D.~Shen, ``Deep chronnectome learning via full
  bidirectional long short-term memory networks for mci diagnosis,'' in {\em
  International conference on medical image computing and computer-assisted
  intervention}, pp.~249--257, Springer, 2018.

\bibitem{dvornek2017identifying}
N.~C. Dvornek, P.~Ventola, K.~A. Pelphrey, and J.~S. Duncan, ``Identifying
  autism from resting-state fmri using long short-term memory networks,'' in
  {\em International Workshop on Machine Learning in Medical Imaging},
  pp.~362--370, Springer, 2017.

\bibitem{rashid2016classification}
B.~Rashid, M.~R. Arbabshirani, E.~Damaraju, M.~S. Cetin, R.~Miller, G.~D.
  Pearlson, and V.~D. Calhoun, ``Classification of schizophrenia and bipolar
  patients using static and dynamic resting-state fmri brain connectivity,''
  {\em Neuroimage}, vol.~134, pp.~645--657, 2016.

\bibitem{el2019hybrid}
A.~El-Gazzar, M.~Quaak, L.~Cerliani, P.~Bloem, G.~van Wingen, and R.~M. Thomas,
  ``A hybrid 3dcnn and 3dc-lstm based model for 4d spatio-temporal fmri data:
  An abide autism classification study,'' in {\em OR 2.0 Context-Aware
  Operating Theaters and Machine Learning in Clinical Neuroimaging},
  pp.~95--102, Springer, 2019.

\bibitem{kunert2019extracting}
J.~M. Kunert-Graf, K.~Eschenburg, D.~Galas, J.~N. Kutz, S.~Rane, and B.~W.
  Brunton, ``Extracting reproducible time-resolved resting state networks using
  dynamic mode decomposition,'' {\em Frontiers in computational neuroscience},
  vol.~13, p.~75, 2019.

\bibitem{hutchison2013dynamic}
R.~M. Hutchison, T.~Womelsdorf, E.~A. Allen, P.~A. Bandettini, V.~D. Calhoun,
  M.~Corbetta, S.~Della~Penna, J.~H. Duyn, G.~H. Glover, J.~Gonzalez-Castillo,
  {\em et~al.}, ``Dynamic functional connectivity: promise, issues, and
  interpretations,'' {\em Neuroimage}, vol.~80, pp.~360--378, 2013.

\bibitem{kawahara2017brainnetcnn}
J.~Kawahara, C.~J. Brown, S.~P. Miller, B.~G. Booth, V.~Chau, R.~E. Grunau,
  J.~G. Zwicker, and G.~Hamarneh, ``Brainnetcnn: Convolutional neural networks
  for brain networks; towards predicting neurodevelopment,'' {\em NeuroImage},
  vol.~146, pp.~1038--1049, 2017.

\bibitem{xingjian2015convolutional}
S.~Xingjian, Z.~Chen, H.~Wang, D.-Y. Yeung, W.-K. Wong, and W.-c. Woo,
  ``Convolutional lstm network: A machine learning approach for precipitation
  nowcasting,'' in {\em Advances in neural information processing systems},
  pp.~802--810, 2015.

\bibitem{li2018brain}
H.~Li and Y.~Fan, ``Brain decoding from functional mri using long short-term
  memory recurrent neural networks,'' in {\em International Conference on
  Medical Image Computing and Computer-Assisted Intervention}, pp.~320--328,
  Springer, 2018.

\bibitem{suresh2008risk}
S.~Suresh {\em et~al.}, ``Risk-sensitive loss functions for sparse
  multi-category classification problems,'' {\em Information Sciences},
  vol.~178, no.~12, pp.~2621--2638, 2008.

\bibitem{jack1994sensory}
C.~R. Jack~Jr {\em et~al.}, ``Sensory motor cortex: correlation of presurgical
  mapping with functional mr imaging and invasive cortical mapping.,'' {\em
  Radiology}, vol.~190, no.~1, pp.~85--92, 1994.

\bibitem{mazaika2009methods}
P.~K. Mazaika, F.~Hoeft, G.~H. Glover, A.~L. Reiss, {\em et~al.}, ``Methods and
  software for fmri analysis of clinical subjects,'' {\em Neuroimage}, vol.~47,
  no.~Suppl 1, p.~S58, 2009.

\bibitem{behzadi2007component}
Y.~Behzadi, K.~Restom, J.~Liau, and T.~T. Liu, ``A component based noise
  correction method (compcor) for bold and perfusion based fmri,'' {\em
  Neuroimage}, vol.~37, no.~1, pp.~90--101, 2007.

\bibitem{craddock2012whole}
R.~C. Craddock {\em et~al.}, ``A whole brain fmri atlas generated via spatially
  constrained spectral clustering,'' {\em Human brain mapping}, vol.~33, no.~8,
  pp.~1914--1928, 2012.

\bibitem{mcauliffe2001medical}
M.~J. McAuliffe, F.~M. Lalonde, D.~McGarry, W.~Gandler, K.~Csaky, and B.~L.
  Trus, ``Medical image processing, analysis and visualization in clinical
  research,'' in {\em Proceedings 14th IEEE Symposium on Computer-Based Medical
  Systems. CBMS 2001}, pp.~381--386, IEEE, 2001.

\end{thebibliography}
\end{document}